\documentclass[12pt,preprint]{aastex}
\begin{document}
\title{Flux upper limit of gamma-ray emission by GRB050713a from MAGIC Telescope observations}

\author{ 
 J.~Albert\altaffilmark{a}, 
 E.~Aliu\altaffilmark{b}, 
 H.~Anderhub\altaffilmark{c}, 
 P.~Antoranz\altaffilmark{d}, 
 A.~Armada\altaffilmark{b}, 
 M.~Asensio\altaffilmark{d}, 
 C.~Baixeras\altaffilmark{e}, 
 J.~A.~Barrio\altaffilmark{d}, 
 M.~Bartelt\altaffilmark{f}, 
 H.~Bartko\altaffilmark{g}
% ...
} 
\altaffiltext{a} {Universit\"at W\"urzburg, Germany} 
\altaffiltext{b} {Institut de F\'\i sica d'Altes Energies, Barcelona, Spain} 
\altaffiltext{c} {Institute for Particle Physics, ETH Zurich, Switzerland} 
\altaffiltext{d} {Universidad Complutense, Madrid, Spain} 
\altaffiltext{e} {Universitat Aut\`onoma de Barcelona, Spain} 
\altaffiltext{f} {Fachbereich Physik, Universit\"at Dortmund, Germany} 
\altaffiltext{g} {Max-Planck-Institut f\"ur Physik, M\"unchen, Germany} 
% ...

\begin{abstract}
The long-duration GRB050713a was observed by the MAGIC Telescope,
40 seconds after the burst onset, and followed up for 37 minutes, until twilight.
The observation, triggered by a SWIFT alert, covered energies above $\approx175\:\mathrm{GeV}$.
Using standard MAGIC analysis, no evidence for a gamma signal was found.
As the redshift of the GRB was not measured directly,
the flux upper limit, estimated by MAGIC, is still compatible
with the assumption of an unbroken power-law spectrum extending
from a few hundred keV to our energy range.
\end{abstract}


\section{Introduction}
Observations of high-energy photons from gamma-ray bursts (GRBs) have much contributed 
to a deeper understanding of their nature. The $\gamma$-ray emission observed by the EGRET detector \citep{Hurley} suggests a power-law spectrum extending up to GeV energies.
This favours an optically thin emission region and a non-thermal origin of the bursts.
As the excessive pair production could be suppressed in the 
presence of relativistic jets \citep{Goodman,Paczynski}, it was concluded 
that relativistic beaming could play an important role for GRBs \citep{Meszaros}.
However, other models also point towards the presence of a strong thermal component in the
GRB spectra \citep{Ryde}.

The observation of  $\gamma$-rays at highest energies is expected to have an important impact
on the modelling of the emission processes, in particular of the early and late afterglow phases of GRBs.
EGRET measurements generally showed the presence of a hard,
long-duration component \citep{Dingus}, consistent with a simple extrapolation
of the MeV spectrum into the high-energy  $\gamma$-ray regime. Recently, an additional,
delayed high-energy component of GRB970417 was found with the TASC detector of EGRET \citep{Gonzalez}.
Several models predict GeV-TeV emission lasting up to the early afterglow \citep{Peer,Dermer}.
Due to the extremely high energies attainable inside relativistic jets,
GRBs are potential sources of Very High Energy (VHE) cosmic rays \cite{Waxman,Vietri},
that can produce in their turn hadronic showers containing VHE $\gamma$-rays.
Other theoretical models predict no emission above few MeV \citep{Lazzati}
or predict strong emission up to GeV, but no emission above 10 GeV \citep{Stern}.
Therefore, measurements at this energy range can be used to test all these competing models.
However, as most of the observed GRBs occur at large redshift, strong attenuation of the 
VHE $\gamma$ flux is expected, as a result of the interaction with low energy photons
of the Metagalactic Radiation Field (MRF) \citep{Nikishov,deJager}.
The knowledge of the redshift is, therefore, important for a precise interpretation.
On the other hand, a detection of VHE $\gamma$-rays provides an indirect
--- and model dependent --- upper limit of its redshift, if some knowledge of the MRF is assumed.

Several observations of GRBs at energies above hundred GeV have been attempted 
\citep{HEGRA,Asgamma}, without showing any indication of a signal.
This is due to a relatively low sensitivity, as in satellite-borne detectors,
and/or a high energy threshold, as in previous generation of Cherenkov telescopes
or in particle detector arrays.  Up to now, only upper limits on the prompt
or delayed emission of GRBs were set by Whipple \citep{whipple}, MILAGRO 
\citep[see][and refs.\ therein]{milagro}, and STACEE \citep{STACEE}.
STACEE, in the same energy region as attainable by MAGIC, was able to follow
GRB050607 after $3\arcmin11\arcsec$ for 1150~s and set an upper limit of its flux as
$\Phi(>100\:\mathrm{GeV})<4.1\times10^{-9}\:\mathrm{cm}^{-2}\,\mathrm{s}^{-1}\approx6$~C.U.\ (\emph{Crab Units}:
$1.5\times10^{-6}\times E(\mathrm{GeV})^{-2.58}\:\mathrm{ph}\cdot\mathrm{cm}^{-2}\,\mathrm{s}^{-1}\mathrm{GeV}^{-1}$).

The situation may change with the new generation of Cherenkov telescopes, 
which achieve a better flux sensitivity and a lower energy threshold.
Nevertheless, as their small field of view does allow prompt observations
only by serendipitous detection, they have to rely on an \emph{external triggering},
as the one provided by the automated satellite link to the 
\emph{Gamma-Ray Burst Coordinates Network} (GCN) \footnote{See \emph{http://gcn.gsfc.nasa.gov/}.},
which broadcasts the coordinates of events triggered and selected by 
dedicated satellite detectors.

Among the new Cherenkov telescopes, MAGIC \citep{MAGIC} is best suited 
for the detection of the prompt emission of GRBs, due to its low energy threshold, 
large effective area, and, in particular, its capability for fast slewing \citep{drive}.
The low trigger threshold, currently $50\:\mathrm{GeV}$ at zenith,
should allow the observation of GRBs even at large redshift,
as lower energy radiation can effectively reach the Earth without interacting much with the MRF.  
Moreover, in its fast slewing mode, MAGIC can be repositioned in $\lesssim 30\:\mathrm{s}$ to any position on the sky: in case of a \emph{Target of Opportunity\/} alert by GCN, an automated procedure 
takes only few seconds to terminate any pending observation, validate the incoming signal 
and start slewing toward the GRB position.
Extrapolating BATSE observed GRB spectra to VHE with an unbroken power-law of power index from the BATSE catalogue, MAGIC is predicted to detect about one GRB per year
at a $5\sigma$ level \citep{Galante}.

In this letter, we report on the analysis of data collected on GRB050713a during 
its prompt emission phase and for the following 37 minutes.

\section{MAGIC observation}
On 2005 July 13 at 4:29:02~UT the BAT instrument on board SWIFT detected a burst 
located at RA~$21^\mathrm{h}22^\mathrm{m}09^{\,}\fs53$ DEC~$+77\degr04\arcmin29^{\,}\farcs50\pm3\arcmin$ \citep{swift}.
The MAGIC alert system received and validated the alert 12~s after the burst,
and data taking started 40~s after the burst original time $(T_0)$ \citep{MAGIC_GCN}.

\begin{figure}
\plotone{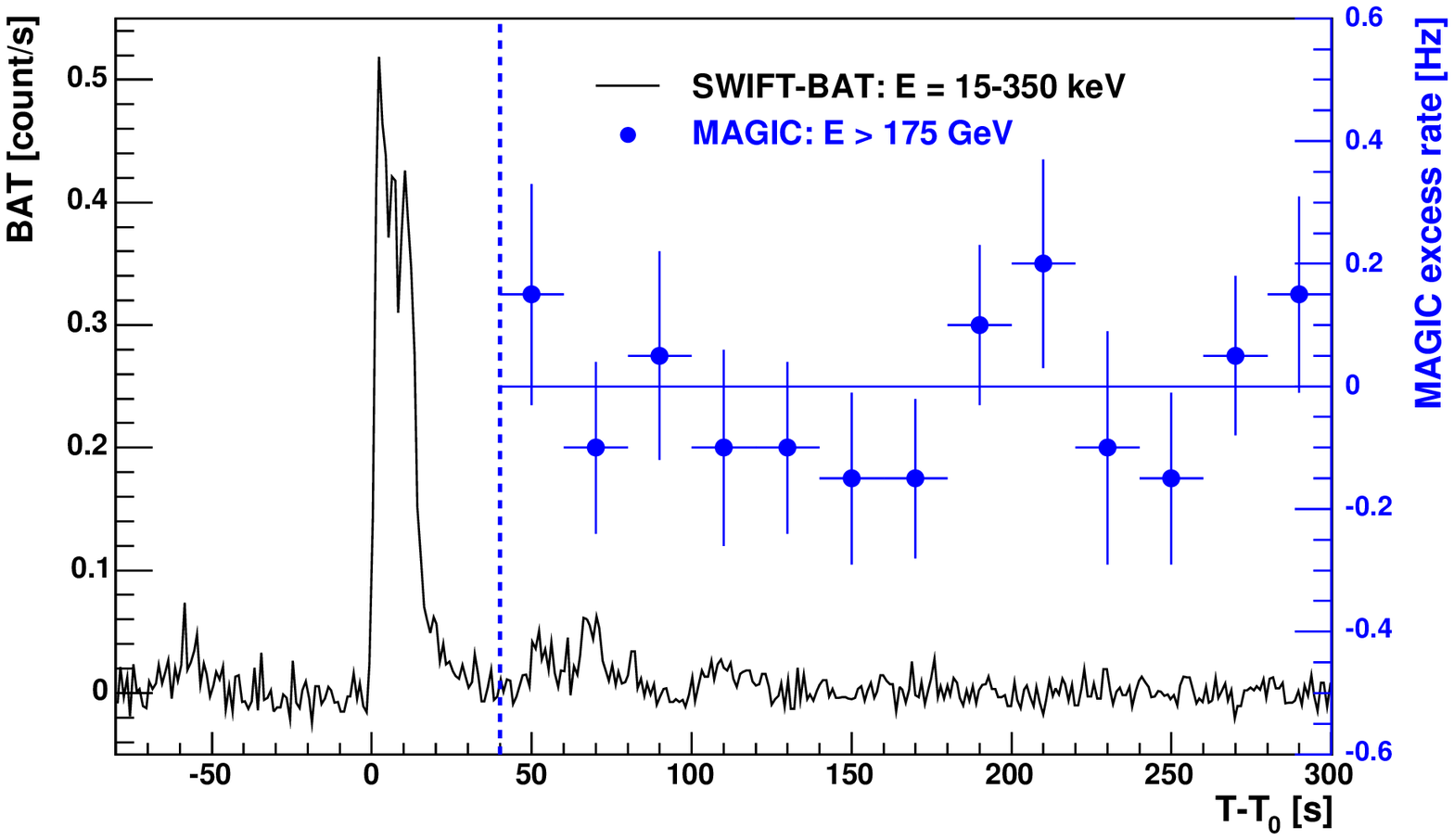}
\caption{\label{fig_ratesswift} MAGIC excess event rate compared with SWIFT-BAT observations.
The vertical line indicates the start of observations with the MAGIC telescope: the prominent
peak seen by SWIFT-BAT occurred before MAGIC observations started.}
\end{figure}

The burst was classified by SWIFT as a bright burst with a duration of $T_{90}=70\pm10\:\mathrm{s}$.
The brightest part of the keV emission occured within $T_0+20$~s,
three smaller peaks followed at $T_0+50$~s, $T_0+65$~s and $T_0+105$~s,
while a \emph{pre-burst\/} peak took place at $T_0-60$~s.
(see figure~\ref{fig_ratesswift}).
The spectrum, over the interval from $T_0-70$~s to $T_0+121$~s, can be fitted
with a power-law with photon index $-1.58\pm0.07$ and yields a fluence of
$9.1\times 10^{-6}\:\mathrm{erg}\cdot\mathrm{cm}^{-2}$
in the $15\div350\:\mathrm{keV}$ range \citep{swift2}. 
The burst triggered also Konus-Wind \cite{Konus}, which measured the spectrum of the burst 
during the first 16~s, that is the duration of the first big peak as 
reported by SWIFT.

\subsection{Data set and analysis}\label{data}
In the local coordinate system of MAGIC, GRB050713a was located at an azimuth angle 
of $-6\degr$ (near North) and a zenith angle of $50\degr$.  The sky region of
the burst was observed for $37\:\mathrm{min}$, until twilight (ON data).
Between $T_0+665\:\mathrm{s}$ and $T_0+686\:\mathrm{s}$,
data taking was interrupted for technical reasons.
A total amount of 258250 atmospheric showers, mainly background, were recorded.
In order to evaluate the contamination of background in the data, the GRB
position was observed again 48~hours later (the so-called \emph{OFF data}).

Data were analyzed using the MAGIC standard software \citep{mars,callisto}.
For optimizing $\gamma$/hadron separation, we simulated $10^5$ $\gamma$ events with
zenith angle ranging between $47\degr$ and $52\degr$, 
energy greater than $10\:\mathrm{GeV}$ and an energy distribution following a power-law spectrum
of index $\beta_{\gamma}=-2.6$. This sample was analyzed in the same way as the data, and was
used for the calculation of the collection area, the sensitivity, and for the energy estimation.
After applying all selection criteria, the sample peaked at around $250\:\mathrm{GeV}$,
which we define as our telescope threshold at this zenith angle.

Data are processed using the standard Hillas analysis \citep{Hillas,Fegan}.
Gamma/hadron separation is performed by means of \emph{Random Forest\/} (RF)\citep{Breiman}, 
a classification method that combines several parameters describing the shape of the image
into a new parameter called \emph{hadronness}, the final $\gamma$/hadron discriminator in our analysis.
The simulated sample was used to optimise, as a function of energy, the cuts in hadronness.
Also the energy of the $\gamma$ was estimated using a RF approach,
yielding a resolution of $\approx30\%$ at $200\:\mathrm{GeV}$.

\begin{figure}
\plotone{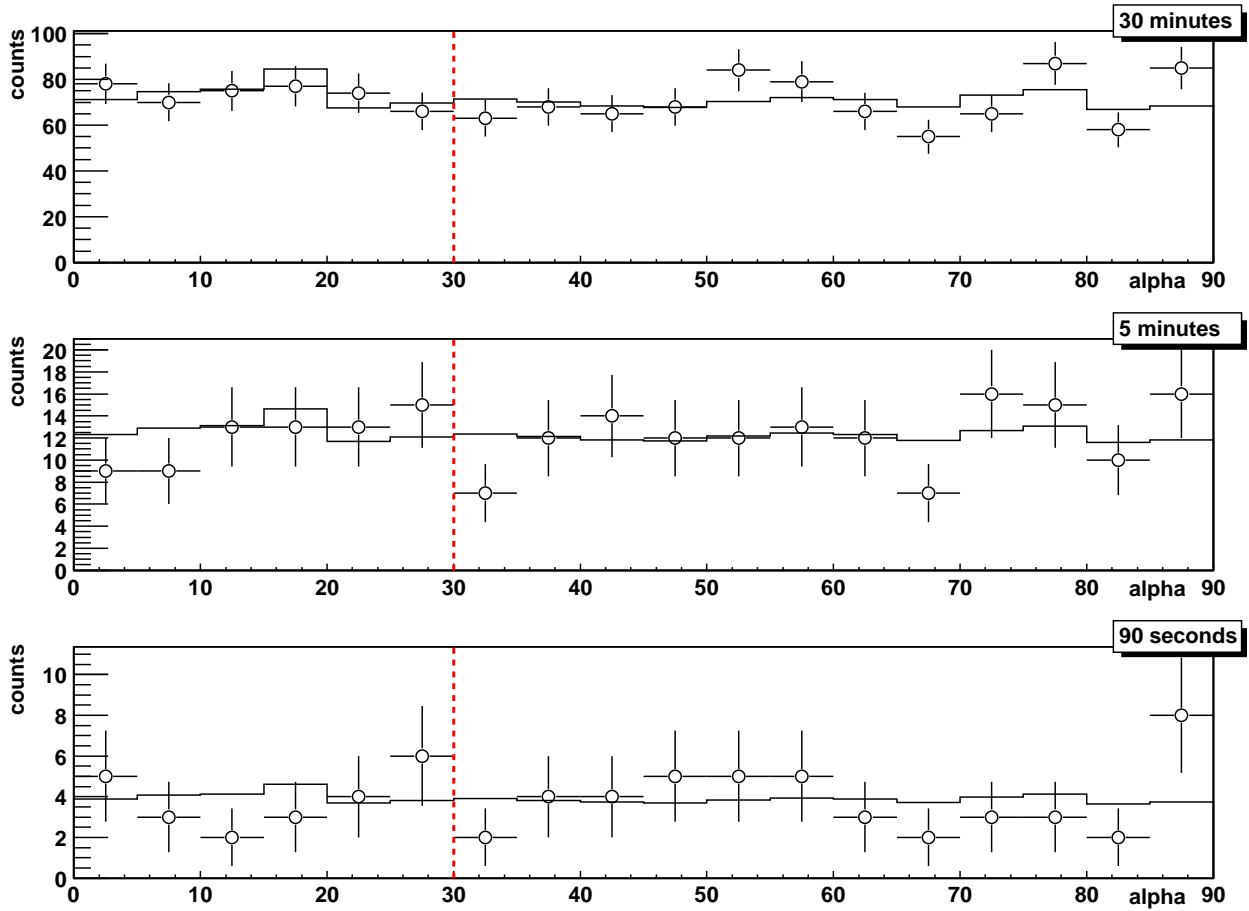}
\caption{\label{fig_alpha}Alpha distributions of events with
$175\:\mathrm{GeV}<E<225\:\mathrm{GeV}$ for three different time intervals 
starting at $T=T_0+40\:\mathrm{s}$: 30~min (top), 5~min (middle), and $90\:\mathrm{s}$ (bottom).
Dots refer to ON data, the line to OFF data.
The vertical line bounds the region where we expect the $\gamma$ signal.}
\end{figure}

The parameter \emph{alpha\/} of the Hillas analysis,
related with the direction of the incoming shower,
is used to evaluate the significance of a signal.
If the telescope is directed at a point-like $\gamma$ source, as a GRB is expected to be,
the alpha distribution of collected photons should peak at $0\degr$,
while it is uniform for isotropic background showers.
According to simulations the $\gamma$-signal at low energies may spread 
in a region defined conservatively by $\textrm{alpha}<30\degr$.
Figure~\ref{fig_alpha} shows the alpha 
distributions for the GRB050713a and for OFF data, divided into three subsets of time 
covering 90~s, 5~min and 30~min, respectively.
No evidence of excess in the signal region is seen.

\subsection{Time analysis}\label{timing}
\begin{figure}
\plotone{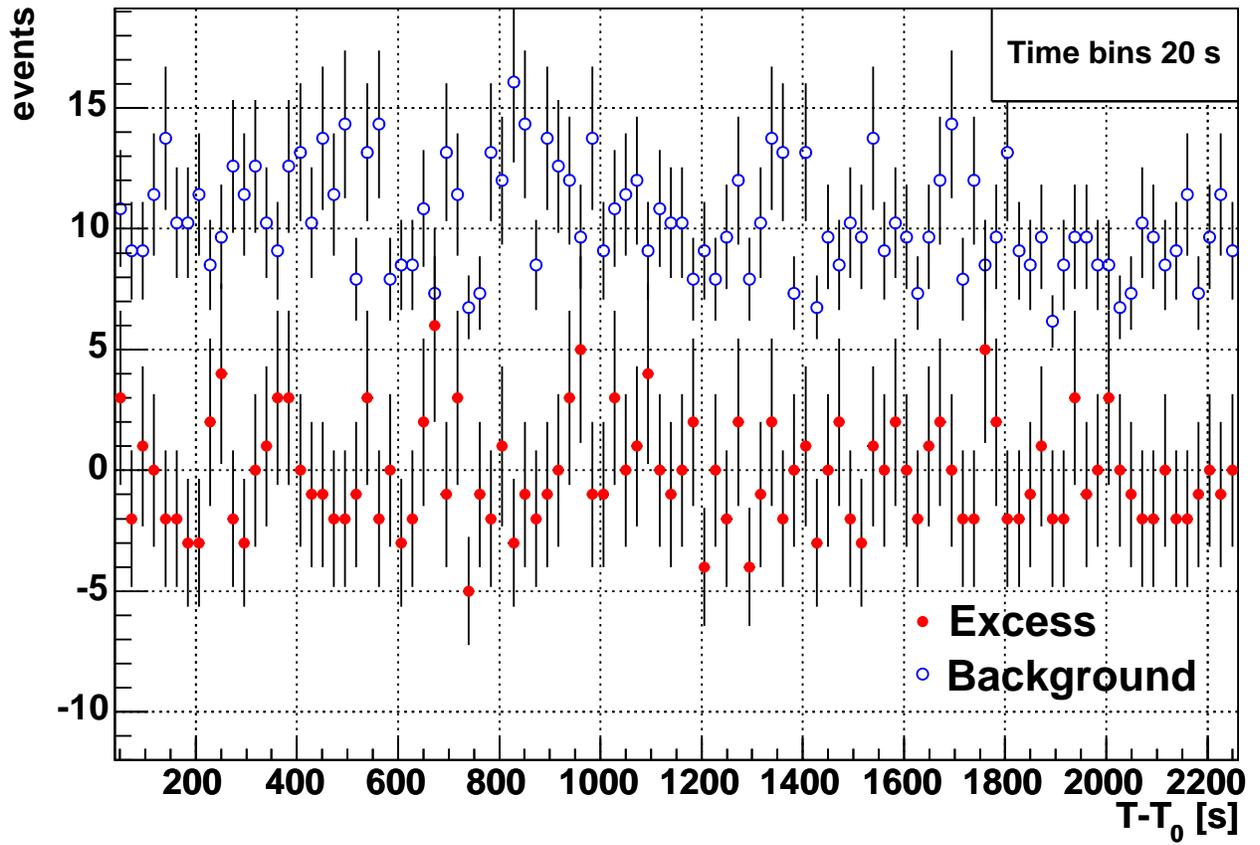}
\caption{\label{fig_rates} Full circles: number of excess events for 20~s intervals,
in the 37~min window after the burst.
Open circles: number of background events in the signal region.}
\end{figure}

A second analysis searching for short time variable $\gamma$-ray signals from GRB050713a has been performed in the range $175\:\mathrm{GeV}<E<225\:\mathrm{GeV}$.
Figure \ref{fig_rates} shows the number of excess events during the first 37 minutes after the burst,
in intervals of 20~s.  The number of expected background events in the signal region (open circles),
estimated from the number of events in the region with $\mathrm{alpha}>30\degr$,
is constant indicating stable experimental condition. The number of excess events is stable
and compatible with statistical fluctuations of the background. The same analysis was applied to the OFF data, with similar results.

\subsection{Flux Upper Limits}
Analysing the data collected during the prompt emission of GRB050713a between $T_0+40$~s and $T_0+130$~s,
we can set upper limits on its flux at 95\% confidence level \citep[see details in:][]{Rolke}.

The upper limit can be used to constrain the prompt emission of the GRB in the VHE range.
Since the observed spectrum is the convolution of the intrinsic spectrum and the
MRF absorption, the limits on the former are thus necessarily model dependent. 

\begin{figure}
\plotone{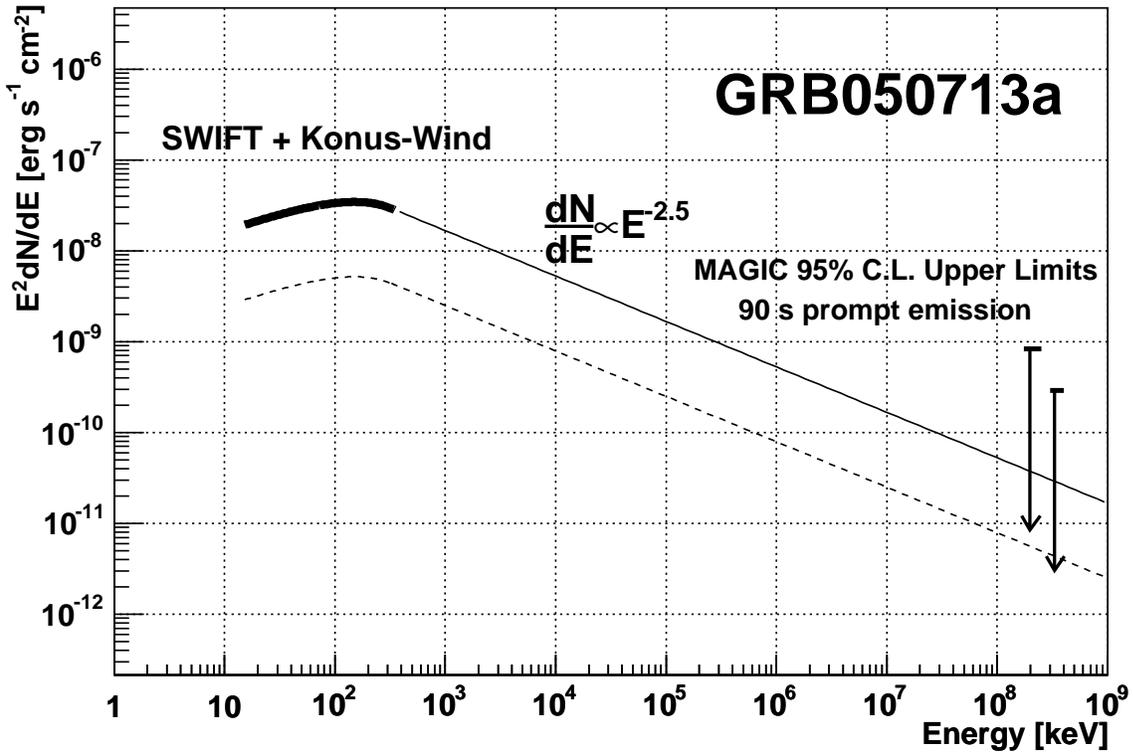}
\caption{\label{fig_flux} Upper limits set by MAGIC on GRB050713a with no redshift correction applied (see text).
 The solid line is the flux measured by SWIFT averaged over the burst $T_{90}$,
 and the energy break is estimated using Konus-Wind data.
 The dashed line represents the fraction of the flux emitted
 between $T_0+40$~s and $T_0+130$~s.}
\end{figure}

First of all, we assumed the GRB spectrum extends to GeV energies
following the Band function \citep{Band}: after the energy break,
estimated by Konus-Wind to be at $\sim355\:\mathrm{keV}$,
the flux follows a power-law of spectral index $\beta=-2.5$,
the mean value of the BATSE distribution \citep[see][]{Preece}.
In this hypothesis, we calculated the upper limit
on the average flux in our energy range
during the entire 90~s interval.  These values are summarised in table~\ref{tab_flux},
and the lowest two energy bins are shown in figure~\ref{fig_flux},
together with the spectrum measured at lower energies by SWIFT and Konus-Wind.

It has to be noted, however, that according to BAT data,
only $10\%\div15\%$ of the total burst fluence
in the $100\:\mathrm{keV}$ region was released in the time window of
the MAGIC observations.  This fraction of the flux is plotted in figure~\ref{fig_flux}
using a dashed line. Adopting a semi-empirical model for the cosmologically evolving MRF \cite{kneiske}, we derived unfolded flux upper limits for various redshift values as shown in table~\ref{tab_flux}.

\section{Conclusions}
MAGIC was able to observe part of the prompt emission phase of a GRB as a response to the alert system provided by the SWIFT satellite.  No excess above 175~GeV was detected neither during the prompt emission phase nor during the following 37 minutes.
We derived an upper limit to the $\gamma$-ray flux between 175 and 1000~GeV.
The observation window covered by MAGIC did not contain the first prominent peak detected at keV energies where the SWIFT and Konus-Wind spectra were taken. Upper limits are compatible with any na\"{\i}ve extensions of the power-law spectrum up to hundreds of GeV.

For the first time a Cherenkov telescope is now able to perform direct observations of the prompt emission phase of GRBs.  Although strong absorption of the high-energy $\gamma$-ray flux by the MRF is expected at high redshifts, given its sensitivity to low fluxes and its fast slewing capabilities, the MAGIC telescope is expected to detect about one GRB per year, if the GRB spectra extend to the hundreds of GeV energy domain.

\acknowledgments
\section*{Acknowledgments}
The construction of the MAGIC Telescope was mainly made possible
by the support of the German BMBF and MPG,
the Italian INFN, and the Spanish CICYT, to whom goes our grateful
acknowledgement.
We are grateful for all the hard work done by the GCN team,
especially Dr.\ Scott Barthelmy, and to all the people of the SWIFT
Science Center who kindly provided us with data,
and tools to analyse them.  In particular, we are indebted
with Prof.\ Guido Chincarini, Dr.\ Abe Falcone, and Prof.\ David Burrows
from the SWIFT Collaboration.
We are also grateful to Dr.\ Nicola Omodei for fruitful discussions on
the physics of GRBs.
We would also like to thank the IAC for the excellent working
conditions at the Observatorio del Roque de los Muchachos in La Palma.
This work was further supported by ETH Research Grant TH~34/04~3
and the Polish MNiI Grant 1P03D01028.

Facilities: \facility{MAGIC}

\clearpage

\clearpage

\clearpage

\clearpage

\clearpage
\begin{table}[tbh]
\begin{center}
\begin{small}
\begin{tabular}{ccccccc}
 \textbf{Energy}
  & \textbf{Excess evts.}
  & \textbf{Eff.\ Area}
  & \multicolumn{4}{c}{\textbf{Flux upper limit}
                       $\times 10^{-9}\left[\mathrm{erg}\:\mathrm{cm}^{-2}\:\mathrm{s}^{-1}\right]$}\\
 $\left[\textrm{GeV}\right]$
  & \textbf{upper limit}
  & $\times 10^8\left[\mathrm{cm}^2\right]$
  & $z=0$ & $z=0.2$ & $z=0.6$ & $z=1$ \\
 $175-225$  & $\phantom08.5$ & 1.7 & $0.83$ (7.6 C.U.) & 1.16 & $\phantom03.42$ & $\phantom010.49$ \\
 $225-300$  &          10.4  & 3.4 & $0.45$ (4.8 C.U.) & 1.07 & $\phantom04.63$ & $\phantom019.32$ \\
 $300-400$  & $\phantom06.0$ & 5.3 & $0.37$ (3.8 C.U.) & 1.35 &          13.20  & $\phantom095.45$ \\
 $400-1000$ & $\phantom04.3$ & 6.5 & $0.13$ (3.3 C.U.) & 0.68 &          25.11  &          293.18  \\
\end{tabular}
\end{small}
\end{center}
\caption{\label{tab_flux} MAGIC upper limit (95\% CL) on GRB050713a
between $T_0+40$~s and $T_0+130$~s (see text).
Limits include a systematic uncertainty of 30\% and have been corrected for the photon absorption by the EBL for different redshift values $z$. For $z=0$ (no correction applied) the flux upper limits are expressed also in Crab Units (C.U.).}
\end{table}


\begin{thebibliography}{}
\bibitem[Atkins et al.\ 2005]{milagro} Atkins, R., et al.\ 2005, \apj, 630, 996
\bibitem[Band et al.\ 1993]{Band} Band, D., et al.\ 1993, \apj, 413, 281
\bibitem[Breiman 2001]{Breiman} Breiman, L.\ 2001, Machine Learning, 45, 5
\bibitem[Bretz et al.\ 2003]{drive} Bretz, T., Dorner, D., \& Wagner, R.\ (MAGIC Coll.) 2003, in Procs.\ $28^\mathrm{th}$ ICRC, Tsukuba (Japan)
\bibitem[Bretz et al.\ 2005]{mars} Bretz, T.\ (MAGIC Coll.) 2005, in Procs.\ $29^\mathrm{th}$ ICRC, Pune (India)
\bibitem[Connaughton et al.\ 1997]{whipple} Connaughton, V., et al.\ 1997, \apj, 479, 859
\bibitem[de Jager \& Stecker 2002]{deJager} de Jager, O.~C., \& Stecker, F.~W.\ 2002, \apj, 566, 738
\bibitem[Dingus 1995]{Dingus} Dingus, B.~L.\ 1995, \apss, 231, 187
\bibitem[Dermer \& Atoyan 2004]{Dermer} Dermer, C. D., \& Atoyan, A.\  2004, A\&A, 418, L5-L8
\bibitem[Fegan et al.\ 1997]{Fegan} Fegan, D.J.\ 1997, J.\ Phys.\ G, 23, 1013
\bibitem[Galante et al.\ 2003]{Galante} Galante, N., Bastieri, D., Gaug, M., Garczarczyk, M., \& Peruzzo, L.\ (MAGIC Coll.) 2003, in Procs.\ $28^\mathrm{th}$ ICRC, Tsukuba (Japan)
\bibitem[Gaug et al.\ 2005]{callisto} Gaug, M., Bartko, H., Cortina, J., Rico, J.\ (MAGIC Coll.) 2005, in Procs.\ $29^\mathrm{th}$ ICRC, Pune (India)
\bibitem[Gonzalez et al.\ 2003]{Gonzalez} Gonzalez, M.~M., Dingus, B.~L., Kaneko, Y., Preece, R.~D., Dermer, C.~D., \& Briggs, M.~S.\ 2003, \nat, 424, 749
\bibitem[Goodman 1986]{Goodman} Goodman, J.\ 1986, \apj, 308, L47
\bibitem[Goetting et al.\ 2003: GCN \#1007 ]{HEGRA} GCN 1007: N.~Goetting et al., 2003
\bibitem[Falcone et al.\ 2005: GCN \#3581]{swift} GCN 3581: A.~Falcone et al., 2005
\bibitem[Palmer et al.\ 2005: GCN \#3597]{swift2} GCN 3597: D.~Palmer et al., 2005
\bibitem[Golenetskii et al.\ 2005: GCN \#3619]{Konus} GCN 3619: S.~Golenetskii et al., 2005
\bibitem[Galante et al.\ 2005: GCN \#3747]{MAGIC_GCN} GCN 3747: N.~Galante et al., 2005
\bibitem[Hillas 1985]{Hillas} Hillas, A.~M.\ 1985, in Proc.\ $19^\mathrm{th}$ ICRC, La Jolla, USA, vol.\ 3, 445
\bibitem[Hurley et al.\ 1994]{Hurley} Hurley, K., et al.\ 1994, \nat, 372, 652
\bibitem[Jarvis et al.\ 2005]{STACEE} Jarvis, B.\ (STACEE Coll.) 2005, in Procs.\ $29^\mathrm{th}$ ICRC, Pune (India)
\bibitem[Kneiske et al.\ 2004]{kneiske} Kneiske, T.~M., Bretz, T., Mannheim, K., Hartmann, D.~H.\ 2004, \aap, 413, 807
\bibitem[Lazzati et al. 2004]{Lazzati} Lazzati, D., Rossi, E., Ghisellini, G., Rees, M.\ 2004, MNRAS, 347, L1
\bibitem[M{\'e}sz{\'a}ros \& Rees 1993]{Meszaros} M{\'e}sz{\'a}ros, P.\, \& Rees, M.J.\ 1993, \apj, 418, L59
\bibitem[Mirzoyan et al.\ 2005]{MAGIC} Mirzoyan, R.\ (MAGIC Coll.) 2005, in Procs.\ $29^\mathrm{th}$ ICRC, Pune (India)
\bibitem[Nikishov 1961]{Nikishov}Nikishov, A.I.\ 1961, Zh.\ Eksp.\ Teor.\ Fiz., 41, 549 (English transl.\ in Soviet
Phys.-JETP Lett., 14, 392 [1962])
\bibitem[Paczynski 1986]{Paczynski} Paczynski, B.\ 1986, \apj, 308, L43
\bibitem[Pe'er \& Waxman 2004]{Peer} Pe'er, A.\, \& Waxman, E.\ 2004, \apj, 603, L1
\bibitem[Preece et al.\ 2000]{Preece} Preece, R.~D., Briggs, M.~S., Mallozzi, R.~S., Pendleton, G.~N., Paciesas, W.~S., \& Band, D.~L.\ 2000, \apjs, 126, 19
\bibitem[Ryde\ 2004]{Ryde} Ryde, F.\  2004, \apj, 614, 827
\bibitem[Rolke{,} L{\'o}pez{,} \& Conrad 2005]{Rolke} Rolke, W., L{\'o}pez, A., \& Conrad, J.\ 2005, Nucl.\ Instr.\ \& Meth.\ A, 551, 493
\bibitem[Stern \& Poutanen\ 2004]{Stern} Stern, B., Poutanen, J.\ 2004, MNRAS, 352, L35-L39
\bibitem[Vietri\ 1995]{Vietri} Vietri, M.\ 1995, \apj, 453, 883
\bibitem[Waxman\ 1995]{Waxman} Waxman, E.\ 1995, Phys. Rev. Lett., 75, 386
\bibitem[Zhou et al.\ 2003]{Asgamma} Zhou, X., et al.\ 2003, in Procs.\ $28^\mathrm{th}$ ICRC, Tsukuba (Japan)
\end{thebibliography}
\end{document}